\documentclass[12pt]{article}
\input{psfig}
\begin{document}
\hfill HU-EP-03/16

\hfill hep-th/0304209\\
 \begin{center}
{\bf\Large On (Orientifold of) type IIA on a Compact
Calabi-Yau}
\vskip 0.1in
A. Misra\footnote{e-mail:misra@physik.hu-berlin.de}\\
Humboldt-Universit\"{a}t zu Berlin\\ Mathematisch-Naturwissenschaftliche Fakult\"{a}t I\\ 
Institut f\"{u}r Physik, Newtonstra\ss e 15, 12489 Berlin-Adlershof, Germany
\end{center} 
\vskip 0.3 true in
\begin{abstract}
We study the gauged sigma model and its mirror Landau-Ginsburg model corresponding to 
type IIA on the  Fermat degree-24 hypersurface in
${\bf WCP}^4[1,1,2,8,12]$ (whose blow-up gives the smooth $CY_3(3,243)$) 
away from the orbifold singularities, and its orientifold by a freely-acting
antiholomorphic involution.
We derive  the Picard-Fuchs equation obeyed by the period integral as defined
in \cite{C,HV}, of the parent ${\cal N}=2$ type IIA theory of \cite{KV} We obtain
the Meijer's basis of solutions to the equation in the large {\it and}
 small complex structure limits (on the mirror Landau-Ginsburg side)
of the abovementioned Calabi-Yau, and make some remarks about the monodromy properties 
associated based on \cite{Morrison}, at the same and another MATHEMATICAlly interesting
point.  Based on a recently shown ${\cal N}=1$ four-dimensional triality \cite{tri1}
between Heterotic on the self-mirror Calabi-Yau $CY_3(11,11)$, 
$M$ theory on ${CY_3(3,243)\times S^1\over{\bf Z}_2}$ and 
$F$-theory on an elliptically fibered $CY_4$ with the base given by ${\bf CP}^1\times$
Enriques surface, we first give a heuristic argument 
that there can be no superpotential generated in the orientifold of 
of $CY_3(3,243)$,
 and then explicitly verify the same using mirror symmetry formulation of \cite{HV}
for the abovementioned hypersurface away from its orbifold singularities. We 
then discuss briefly the sigma model and the mirror Landau-Ginsburg model
corresponding to the resolved Calabi-Yau as well.
\end{abstract}
\clearpage

\section{Introduction}

The periods are the building blocks, e.g., for getting the prepotential
in ${\cal N}=2$ type II theories compactified on a Calabi-Yau that in turn
determines the gauge and the Yukawa couplings, as well as getting the
superpotential in the context of ${\cal N}=1$ type II theories on
(orientifolds of) Calabi-Yau with(out) branes/fluxes. It is in this regard
that the Picard-Fuchs equation satisfied by the periods, become quite 
important.
In the context of toric geometry, the Picard-Fuchs equation has been obtained in the past
essentially based on two approaches 
- one based on  \cite{Griffiths}, and the other based on \cite{GZK}
followed in \cite{Klemmetal}. 
It is \cite{Klemmetal} that will be somewhat closer in spirit to the way we derive the solution
to the Picard-Fuchs equation for the degree-24  Fermat hypersurface in ${\bf WCP}^4[1,1,2,8,12]$
corresponding to the compact $CY_3(3,243)$, away from its orbifold singularities. 
However, in the literature, it is largely
the large complex structure limit in the moduli space that has been
considered where one has to either actually evaluate the integral
\footnote{The Griffiths definition of the period (\cite{Candetal})
for, e.g., one-paramater hypersurface $p(x_i;\psi)$ in ${\bf WCP}^4$ is: 
${1\over(2i\pi)^5}\int_{\gamma_1\times...\gamma_4}
{x_5\prod_{i=1}^4dx_i\over p(x_i;\psi)}|_{x_5=1}$
=${1\over(2\pi i)^5}\int_{\gamma_1\times...
\gamma_5}{\prod_{i=1}^5dx_i\over p(x_i;\psi)}$ where 
$\gamma_i:|x_i|=\delta$. The integral is then evaluated in the
limit:$\psi\rightarrow\infty$. We however do not
restrict ourselves to  this limit (besides, we are largely
interested in Fermat hypersurfaces as the $p$), and hence we
do not wish to directly evaluate the integral.}
or solve the indicial equation
to get the ``fundamental period" and then generate other (logarithmic) 
solutions from it by the  action of derivatives. Part of the motivation for 
this
work is to follow the alternative formulation of \cite{HV} for 
deriving the Picard-Fuchs
equation for a definition of period integral more suited to evaluating BPS mass formulae,
and to obtain solutions valid in the large {\it and} small complex structure limit.
The reason for studying $CY_3(3,243)$ is not arbitrary. It stems from the following.
In \cite{tri1}, we had obtained at the level of spectrum matching, an
${\cal N}=1$ triality between Heterotic on the self-mirror Voisin-Borcea Calabi-Yau
$CY_3(11,11)$, $M$ theory on the `barely $G_2$-manifold (i.e., with $SU(3)\times{\bf Z}_2$
holonomy) constructed from a freely acting antiholomorphic involution on $CY_3(3,243)\times S^1$,
 and $F$ theory on an elliptically fibered 4-fold whose base was ${\bf CP}^1\times$
Enriques surface\footnote{There was an apparent puzzle raised in \cite{tri1} on the $F$-theory
side - the Hodge data of the expected 4-fold does not match the Hodge data of the 
freely acting orbifold of $CY(3,243)\times T^2$ that one would naively have guessed on the 
basis of known dualities between type $IIA$, Heterotic and (and definition of) $F$ theory. Of
course, the 4-fold with the derived Hodge data and fibration structure, has to exist, because
the $F$-theory dual has to exist.}.
We would like to be able to strengthen this triality by matching interesting calculable quantities
on all sides. In this paper, we take a small step in that direction by first heuristically
arguring a null superpotential on the $F$-theory and Heterotic sides, and then explicitly 
verifying the same in type $IIA$, though for this work, away from the orbifold singularities
of the Fermat hypersurface in ${\bf WCP}^4[1,1,2,8,12]$.

Hence, to summarize, we wish to study the compact $CY_3(3,243)$ as well as a particular
freely acting involution thereof (using which one can get the ${\cal N}=1$ type $IIA$ 
background of \cite{VW}, and its $M$-theory uplift of \cite{tri1}), 
from the point of view gauged linear sigma model (and mirror symmetry).

The plan of the  rest of the paper is as follows. In section {\bf 2}, we describe the 
basics relevant to gauged linear sigma model and its mirror Landau-Ginsburg model, and set up
the definition of the period integral for compact Calabi-Yau manifolds, using \cite{HV}.
In section {\bf 3}, we derive the Picard-Fuchs equation as well as study the Meijer
basis of solutions in the large and small complex structure limits (on the 
mirror Landau-Ginsburg side) of the Calabi-Yau. 
We also study the Picard-Fuchs equation at a MATHEMATICAlly interesting point,
which in terms of rescaled coordinates, corresponds to the usually troublesome
$z=1$ point. In section {\bf 4},
we give a plausibility argument, to begin with, to expect a null superpotential in the
aforemetioned free orientifold of type $IIA$, and then explicitly check the 
same using mirror symmetry
arguments of \cite{AAHV}. Section {\bf 5} has the conclusion as well as future directions including a brief discussion on the gauged linear sigma model and
the mirror Landau-Ginsburg model corresponding to the resolved $CY_3(3,243)$.

\section{The Gauged Linear Sigma Model and the mirror Landau Ginsburg Model}

Consider the Calabi-Yau 3-fold given as a degree-24 Fermat 
hypersurface in the weighted projective
space ${\bf WCP}^4[1,1,2,8,12]$:
\begin{equation}
\label{eq:Fermat}
P=z_1^{24}+z_2^{24}+z_3^{12}+z_4^{3}+z_5^2=0.
\end{equation}
This is transverse, as $P={\partial P\over\partial z_{i=1,2,3,4,5}}=0$ has only
$(0,0,0,0,0)$ as the solution, which does not belong to the hypersurface. However, $P$
has a ${\bf Z}_2$-singularity curve and a ${\bf Z}_4$-singularity point. To see this,
in the defintion:
\begin{equation}
\label{eq:def}
(z_1,z_2,z_3,z_4,z_5)\sim(\lambda z_1,\lambda z_2,\lambda^2 z_3,\lambda^8 z_4,\lambda^{12}
z_5), \lambda\in{\bf C}^*,
\end{equation}
set $\lambda=-1$ to see that one gets a ${\bf Z}_2$-singularity along $(0,0,z_3,z_4,z_5)$
which is a curve in the Calabi-Yau, and if one sets $\lambda=i$, one 
gets a ${\bf Z}_4$-singularity along the point $(0,0,0,z_4,z_5)$. The singularity
resolution can be summarized by the following. The ${\bf Z}_2$-singularity is blown up
by a ${\bf CP}^1$ so that the Calabi-Yau is a $K3$-fibration over this ${\bf CP}^1$,
and the ${\bf Z}_4$ singularity is resolved by taking the $K3$ fiber itself to be
an elliptic firbation over another ${\bf CP}^1$. This gives the smooth $CY_3(3,243)$
\footnote{Using Greene-Plesser's prescription, the mirror $CY_3(243,3)$ is obtained by
${P\over{\bf Z}_6\times{\bf Z}_{12}}$ the generators of the orbifolds being given by:
\begin{eqnarray}
\label{eq:IIBmirror}
& & (z_1,z_2,z_3,z_4,z_5)\rightarrow(z_1,e^{{-2i\pi\over12}}z_2,e^{{2i\pi\over12}}z_3,z_4,z_5)
\nonumber\\
& & (z_1,z_2,z_3,z_4,z_5)\rightarrow(e^{{2i\pi\over24}}z_1,e^{{-2i\pi\over24}}z_2,z_3,z_4,z_5)
\nonumber\\
\end{eqnarray}\nonumber}.
In this paper, except for the last section,
we assume that one is away from the orbifold singularities mentioned above.
In the section on Conclusion and Future directions (section {\bf 5}), we
briefly discuss the gauged linear sigma model and its mirror Landau-Ginsburg
model for the resolved $CY_3(3,243)$.

Type IIA theory on the above Calabi-Yau can be described, based on \cite{Witten}, by
a gauged linear sigma model by six chiral superfields ${\cal X}_i, i=1,...,6$ with $U(1)$ charges
(1,1,2,8,12,-24) satisfying the following constraint:
\begin{equation}
\label{eq:Dterm}
|{\cal X}_1|^2+|{\cal X}_2|^2+2|{\cal X}_3|^3+8|{\cal X}_4|^2+12|{\cal X}_5|^2
-24|{\cal X}_6|^2=r,
\end{equation}
with the linear sigma model superpotential $W$ given by 
\begin{equation}
\label{eq:Wlinear sigma}
W({\cal X}_i)=X_6P({\cal X}_{1,...,5}).
\end{equation}
The mirror Landau-Ginsburg (LG) theory will be given in terms of a vector multiplet with
field strength $F$ and twisted chiral superfields $Y_i, i=1,...,6$, satisfying the
constraint:
\begin{equation}
\label{eq:constTCS}
Y_1+Y_2+2Y_3+8Y_4+12Y_5-24Y_6=t(=r+i\theta),
\end{equation}
with
\begin{equation}
\label{eq:TCSCS}
Re(Y_i)=|{\cal X}_i|^2.
\end{equation}
The mirror LG superpotential $\tilde{W}_{LG}$ is given by
\begin{equation}
\label{eq:WLG1}
\tilde{W}_{LG}=(W_{LG}\equiv)\sum_{i=1}^6e^{-Y_i}+F(\sum_{i=1}^6w_iY_i - t),
\end{equation}
where $w_i, i=1,...,6$ are the weights. The constraint (\ref{eq:constTCS}) can be solved
by:
\begin{equation}
e^{-Y_i}=X_i^{{24\over w_i}}.
\end{equation}
Solving for $Y_6$, one gets:
\begin{equation}
e^{-Y_6}=e^{{t\over24}}e^{-({Y_1\over24}+{Y_2\over24}+{Y_3\over12}+{Y_4\over3}+{Y_5\over2})}
=e^{{t\over24}}X_1X_2X_3X_4X_5.
\end{equation}
Hence,
\begin{equation}
\label{eq:WLG2}
W_{LG}=\sum_{i=1}^5e^{-Y_i}+e^{{t\over24}}e^{-\sum_{i=1}^5{w_i\over24}Y_i}.
\end{equation}
The period integral as defined in \cite{HV} for compact manfiolds is:
\begin{equation}
\label{eq:periodieuf}
\Pi(t)\equiv d(\equiv{\rm dimensionality\ of hypersurface})
{d\over dt}\int dF\prod_{i=1} dY_i e^{-\tilde{W}_{LG}}.
\end{equation}
For the Calabi-Yau under consideration, one thus gets:
\begin{eqnarray}
\label{eq:PeriodCYFermat}
& & \Pi(t)=24 {d\over dt}\int dY_6\prod_{i=1}^5\delta(\sum_{i=1}^5w_iY_i - 24 Y_6 - t) 
e^{-W_{LG}}\nonumber\\
& & =24{(-24)^5\over{1.1.2.8.12}}{d\over dt}\int\prod_{i=1}{dX_i\over X_i}
e^{-\sum_{i=1}^5X_i^{{24\over w_i}}
-e^{t\over24}\prod_{i=1}^5X_i}\nonumber\\
& & =-{(-24)^5\over
1.1.2.8.12}\int\prod_i dX_i e^{-\sum_{i=1}^5X_i^{{24\over w_i}}-e^{{t\over24}}
\prod_{i=1}^5X_i}.
\end{eqnarray}

\section{Picard-Fuchs Equation}

In this section we derive the Picard-Fuchs equation from the gauged linear sigma model on
the Calabi-Yau given by (\ref{eq:Fermat}). 
To obtain the Picard-Fuchs equation, following \cite{HV}, consider:
\begin{equation}
\label{PF1}
\Pi(t,\{\mu_i\})
\equiv\int\prod_{i=1}^5dY_i\delta(\sum_{i=1}^6w_iY_i-t)e^{-\sum_{i=1}^5\mu_ie^{-Y_i}}
\end{equation}
such that:
\begin{equation}
\label{eq:BC}
\Pi(t,\{\mu_i=1\})\equiv\Pi(t).
\end{equation}
Writing:
\begin{equation}
\label{eq:PF2}
\Pi(t)=24{d\over dt}\tilde{\Pi}(t),
\end{equation}
and then defining a corresponding $\tilde{\Pi}(t,\{\mu_i\})$,
one then obtains the following Picard-Fuchs equation
\begin{equation}
\label{eq:PF3}
\prod_{i=1}^5\biggl({\partial\over\partial\mu_i}\biggr)^{w_i}\tilde{\Pi}(t,\{\mu_i\})
=e^{-t}\biggl({\partial\over\partial\mu_6}\biggr)^{-w_6=24}\tilde{\Pi}(t,\{\mu_i\}).
\end{equation}
One notes that (See \cite{HV}):
\begin{equation}
\label{eq:tprimedieuf}
\int\prod_i dY_i\delta(\sum_{i=1}^6w_iY_i - t)e^{-\sum_i\mu_iY_i}\ \stackrel{Y_i\rightarrow Y_i+ln\mu_i}{
\longrightarrow}\int\prod_i dY_i
\delta(\sum_{i=1}^6w_iY_i-t^\prime)e^{-\sum_iY_i},
\end{equation}
where
\begin{equation}
t^\prime=t - ln\biggl({\mu_1\mu_2\mu_3^2\mu_4^8\mu_5^{12}\over\mu_6^{24}}\biggr).
\end{equation}
Hence, one 
sees that:
\begin{equation}
\label{eq:equivPis}
\Pi(t,\{\mu_i\})=\Pi(t^\prime,\{\mu_i=1\}).
\end{equation}
If 
\begin{equation}
\label{eq:Thder1}
{\partial\over\partial\mu_i}\rightarrow{w_i\over\mu_i}{\partial\over\partial t^\prime},
\end{equation}
then
\begin{equation}
\label{eq:Thder2}
{\partial^{M_i}\over\partial\mu_i^{M_i}}|_{{\partial\over\partial t^\prime}
\equiv\Theta}\rightarrow\prod_{j=0}^{M_i-1}(m_j\Theta - j),
\end{equation}
using which one gets:
\begin{eqnarray}
\label{eq:Thder3}
& & {\partial\over\partial\mu_{1,2}}\rightarrow-\Theta,\nonumber\\
& & {\partial^2\over\partial\mu_3^2}\rightarrow-2\Theta(-2\Theta-1),\nonumber\\
& & {\partial^8\over\partial\mu_4^8}\rightarrow\prod_{j=0}^7(-8\Theta-j),\nonumber\\
& & {\partial^{12}\over\partial\mu_5^{12}}\rightarrow\prod_{j=0}^{11}(-12\Theta - j),
\nonumber\\
& & {\partial^{24}\over\partial\mu_6^{24}}\rightarrow\prod_{j=0}^{23}(24\Theta - j).
\end{eqnarray}
Using (\ref{eq:Thder2}), one gets the following Picard-Fuchs equation:
\begin{equation}
\label{eq:PF}
(-\Theta)^2(-2\Theta)(-2\Theta-1)\prod_{j=0}^7(-8\Theta - j)\prod_{j=0}^{11}(-12\Theta - j)
\tilde{\Pi}(t^\prime)
=e^{-t^\prime}\prod_{j=0}^{23}(24\Theta - j)\tilde{\Pi}(t^\prime).
\end{equation}

\subsection{Solution to the Picard-Fuchs Equation}

In this subsection, we discuss the solution to the Picard-Fuchs equation (\ref{eq:PF}).
We will do so around
 (a) $e^{-t^\prime}\equiv z=0,\infty$ which could be interpreted as 
the large and small complex structure limits (on the mirror Landau-Ginsburg
side), respectively, 
and (b) $e^{t^\prime}=2229025112064$ - a ``MATHEMATICA"lly interesting
point.

\subsubsection{Solution around $e^{-t^\prime}=0,\infty$}

One notes that $z\equiv e^{-t^\prime}=e^{-t}{\prod_{i=1}^5\mu_i^{w_i}\over
\mu_6^{\sum_{j=1}^5w_j}}$ can be made small by, e.g., taking
$\mu_6\rightarrow\infty$ corresponding to the large complex structure limit
on the mirror Landau-Ginsburg side.
On the other hand by, e.g., taking $\mu_6\rightarrow0$, corresponding to
the small complex structure limit (on the mirror Landau-Ginsburg side), 
one can make $|z|>>1$. It is usually
the large complex structure limit that has been largely dealt with in the
literature. Additionally, for finite values of the complex structure
deformation parameters $\mu_i$'s,
$|z|<<1$ can alternatively correspond to large size of the Calabi-Yau,
parameterized by (large) $t$.
In this work, we solve for both  the limits with equal ease,
following \cite{GL}. The method has the additional advantage that
the solutions with logarithmic terms corresponding to the large complex
structure limit, do not have to be obtained by any process of differentiation
of solutions with some additional parameter that is eventually set to zero.
One gets them as naturally as the one without the logarithmic terms.

By setting $e^{-t^\prime}\equiv z$ and $\Delta_z\equiv z{d\over dz}$, one sees that the
Picard-Fuchs equation (\ref{eq:PF}) can be written as:
\begin{equation}
\label{eq:PF4}
\Delta^2_z\Delta_z(\Delta_z-{1\over2})\prod_{j=1}^8(\Delta_z-{j-1\over8})\prod_{j=1}^{12}
(\Delta_z-{j-1\over12})\tilde{\Pi}
=z{(24)^24\over(2^2.8^8.12^{12})}\prod_{j=1}^{24}(\Delta_z+{j-1\over24})\tilde{\Pi}.
\end{equation}
The numerical factor on the RHS can be absorbed into $z$ after a suitable rescaling and
noting that $\Delta_z$ remains invariant under such a rescaling.  
Comparing (\ref{eq:PF4} with the following differential equation for a generalized
Hypergeometric function (See \cite{GL})\footnote{We thank C.I.Lazirou for bringing
\cite{GL} to our attention.} 
$
\ _pF_q\left(\begin{array}{cccc}\\
\alpha_1 & \alpha_2 & \alpha_3 & ....\ \alpha_p\\
\beta_1 & \beta_2 & \beta_3 & ....\ \beta_q\\
\end{array}\right)
$:
\begin{equation}
\label{eq:pFqdieuf}
\biggl[
\Delta_z\prod_{i=1}^q(\Delta+\beta_i-1)-z\prod_{j=1}^p(\Delta+\alpha_j)\biggr]\tilde{\Pi}=0,
\end{equation}
one notes:
\begin{eqnarray}
\label{eq:pFqcomp}
& & p=24,\ q=23;\nonumber\\
& & \beta_1=\beta_2=1,\ \beta_3={1\over2},\ \beta_i={9-i\over8}, i=4,...,11,\
\beta_i={13-i\over12}, i=12,...,23;\nonumber\\
& & \alpha_i={i-1\over24}, i=1,...,24.
\end{eqnarray}
One 
solution to (\ref{eq:PF4}) can be written in terms of the following generalized
hypergeometric function
\begin{equation}
\label{eq:PFsol1}
\ _pF_q\left(\begin{array}{ccccccc}\\
0 & {1\over24} & {2\over24} & {3\over24} & {4\over24} & {5\over24} & ....\ {23\over24}\\
1 & 1 & {1\over2} & {5\over8} & ...\ -{2\over8} & {1\over12} & ...\ -{10\over12}\\
\end{array}\right).
\end{equation}
From the above solution, using properties involving the generalized hypergeometric
function $\ _pF_q$ and the Meijer function $I$:
\begin{eqnarray}
\label{eq:IpFqprops}
& & \ _pF_q\left(\begin{array}{cccc}\\
\alpha_1 & \alpha_2 & \alpha_3 & ....\ \alpha_p\\
\beta_1 & \beta_2 & \beta_3 & ....\ \beta_q\\
\end{array}\right)(z)={\prod_{i=1}^p\Gamma(\beta_i)\over\prod_{j=1}^q\Gamma(\alpha_j)}
I\left(\begin{array}{c|c}\\
0 & \alpha_1...\alpha_p\\
&\nonumber\\ \hline
&\nonumber\\ 
. & \beta_1...\beta_q\\
\end{array}\right)(-z)\nonumber\\
& & I\left(\begin{array}{c|c}\\
a_1...a_A & b_1...b_B\\
&\nonumber\\ \hline
&\nonumber\\ 
c_1...c_C & d_1...d_D\\
\end{array}\right)= I\left(\begin{array}{c|c}\\
a_1...1-d_l...a_A & b_1...b_B\\
&\nonumber\\ \hline
&\nonumber\\ 
c_1...c_C & d_1...\hat{d}_l...d_D\\
\end{array}\right)(-z),
\end{eqnarray}
one generates the following additional 23 solutions that together with (\ref{eq:pFqcomp})
forms the Meijer basis of solutions to the Picard-Fuchs equation (\ref{eq:PF4}):
\begin{eqnarray}
\label{eq:Meijerbasis}
& & (a_1)\ I\left(\begin{array}{c|c}\\
00& 0{1\over24}...{23\over24}\\
&\nonumber\\ \hline
&\nonumber\\ 
. & 1{1\over2}{5\over8}...-{2\over8}{1\over12}...-{10\over12}\\
\end{array}\right)(z)\nonumber\\
& & (a_2)\ I\left(\begin{array}{c|c}\\
000 & 0{1\over24}...{23\over24}\\
&\nonumber\\ \hline
&\nonumber\\ 
.&{1\over2}{5\over8}...-{2\over8}{1\over12}...-{10\over12}\\
\end{array}\right)(-z)\nonumber\\
& & (a_3)\ I\left(\begin{array}{c|c}\\
000{1\over2} & 0{1\over24}...{23\over24}\\
&\nonumber\\ \hline
&\nonumber\\ 
.&{5\over8}...-{2\over8}{1\over12}...-{10\over12}\\
\end{array}\right)(z)\nonumber\\
& & (a_4)\ I\left(\begin{array}{c|c}\\
000{1\over2}{3\over8} & 0{1\over24}...{23\over24}\\
&\nonumber\\ \hline
&\nonumber\\ 
. & {4\over8}...-{2\over8}{1\over12}...-{10\over12}\\
\end{array}\right)(-z)\nonumber\\
& & ...................................................................................................................\nonumber\\
& & (a_{11})\ I\left(\begin{array}{c|c}\\
000{1\over2}{3\over8}{4\over8}{5\over8}...{10\over8} & 0{1\over24}...{23\over24}\\
&\nonumber\\ \hline
&\nonumber\\ 
.&{1\over12}...-{10\over12}\\
\end{array}\right)(z)\nonumber\\
& & (a_{12})\ I\left(\begin{array}{c|c}\\
000{1\over2}{3\over8}{4\over8}{5\over8}...{10\over8} {11\over12} & 0{1\over24}...{23\over24}\\
&\nonumber\\ \hline
&\nonumber\\ 
. & 0...-{10\over12}\\
\end{array}\right)(-z)\nonumber\\
& & .......................................................................................................................\nonumber\\
& & (a_{23})\ I\left(\begin{array}{c|c}\\
000{1\over2}{3\over8}{4\over8}{5\over8}...{10\over8} {11\over12} 1{13\over12}...{22\over12}
& 0{1\over24}...{23\over24}\\
&\nonumber\\ \hline
&\nonumber\\ 
. & .\\
\end{array}\right)(z).
\end{eqnarray}
Now, to get an infinite series expansion in $z$ for $|z|<1$ as well as $|z|>1$, 
one uses the following
Mellin-Barnes integral represention for the Meijer's function $I$:
\begin{equation}
I\left(\begin{array}{c|c}\\
a_1...a_A & b_1...b_B\\
& \\ \hline
& \\
c_1...c_C & d_1...d_D\\
\end{array}\right)(z)={1\over2\pi i}\int_\gamma ds {\prod_{i=1}^A\Gamma(a_i-s)\prod_{j=1}^B
\Gamma(b_j+s)\over\prod_{k=1}^C\Gamma(c_k-s)\prod_{l=1}^D\Gamma(d_l+s)}z^s,
\end{equation}
where the contour $\gamma$ lies to the right 
of the poles at $s+b_j=-m\in{\bf Z}^-\cup\{0\}$ and
to the left of the poles at $a_i-s=-m\in{\bf Z}^-\cup\{0\}$ (See Fig 1).
\begin{figure}[htbp]
\vskip -2in
\centerline{{\psfig{file=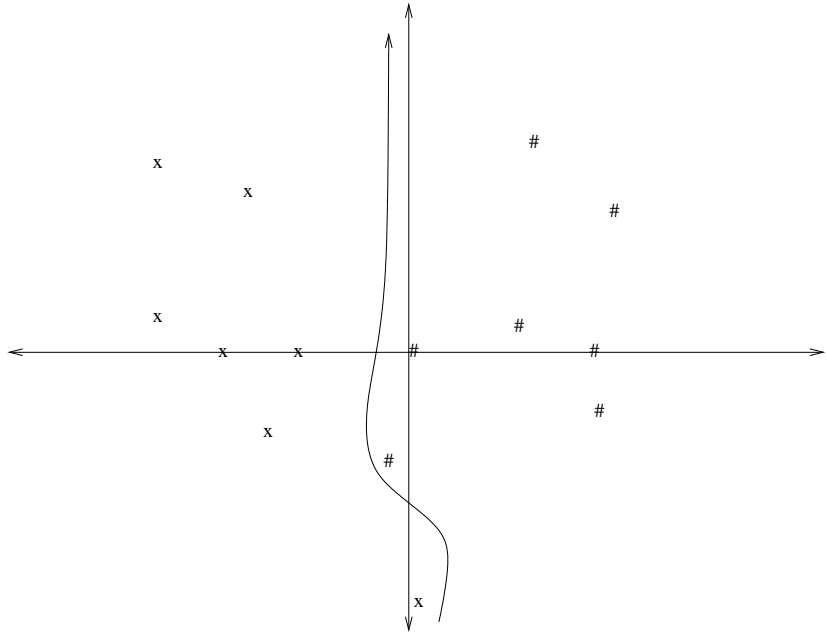,width=0.8\textwidth}}}
\vskip -1in
\caption{The contour $\gamma$ for $I$: The poles $s=a_i+m$ are denoted by $\#$ and the
poles $s=-m-b_j$ are denoted by $x$}
  \end{figure}
Given that one is interested in the region $|z|<1$, one can deform the contour $\gamma$
to $\gamma^\prime$  in Fig 2 below:
\begin{figure}[htbp]
\vskip -2in
\centerline{{\psfig{file=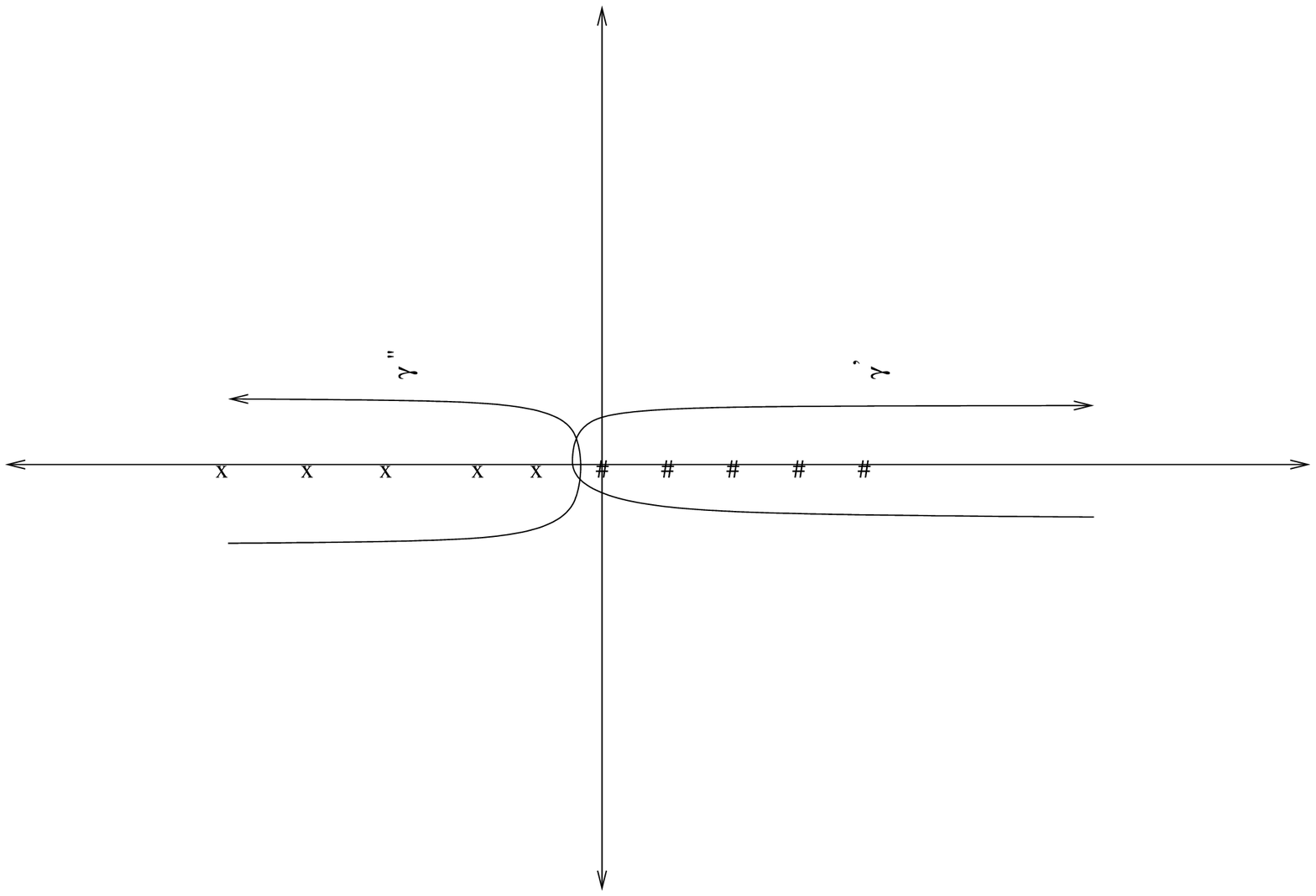
,width=0.8\textwidth}}}
\vskip -1in
\caption{The deformed contour $\gamma^\prime$ valid for $|z|<1$, and $\gamma^{\prime\prime}$ 
valid for $|z|>>1$}
\end{figure}
In the following we evaluate the Mellin-Barnes integral for the Meijer basis of solutions based
on the techniques of \cite{GL}.

(a) 
\begin{eqnarray}
\label{eq:sol1}
& & I\left(\begin{array}{c|c}\\
0 & 0 {1\over24} {2\over24} {3\over24} {4\over24}...{23\over24}\\
& \\ \hline
&\\
. & 1 1 {1\over2} {5\over8}...-{2\over8} {1\over12}...-{10\over12} \\
\end{array}\right)(-z)=\nonumber\\
& & ={1\over2\pi i} \int_{\gamma^\prime}{\Gamma(-s)\Gamma(s)\prod_{j=1}^{23}\Gamma({j\over24}+s)\over
[\Gamma(1+s)]^2\Gamma({1\over2}+s)\prod_{k=-2}^5\Gamma({i\over8}+s)\prod_{l=10}^1\Gamma({l\over12}+s)}
(-z)^s
\end{eqnarray}
Using ${\Gamma(s)\Gamma(-s)\over\Gamma(1+s)^2}={\Gamma(-s)\over s\Gamma(s+1)};\ 
\Gamma(-s)=(-)^{m+1}{\Gamma(-s+m+1)\over(s-m)(s-m+1)_m}$ 
implying a pole of order 2 at $s=0$
and a pole of order 1 at $s=m\in{\bf Z}^+$, and the fact that the pole at $s=0$ will not contribute
to the residue because the residue will involve multiplication of a finite quantity
\footnote{The finite quantity being: 
$\sum_{j=1}^{23}\Psi({j\over24})+ln(-z)+2\gamma-\Psi({1\over2})
-\sum_{k=-2}^5\Psi({k\over8})-\sum_{l=-{10}}^1\Psi({l\over12})$.} with
${\prod_{j=1}^{23}\Gamma({j\over24})\over\Gamma({1\over2})\prod_{k=-2}^5\Gamma({k\over8})\prod_{l=10}^2
\Gamma({l\over12})}$ implying that there will be a ${1\over\Gamma(0)^2}$ factor nullifying the contribution
to the residue
Hence, the final answer is:
\begin{eqnarray}
\label{eq:sol1fin}
& & \sum_{m=1}^\infty{\prod_{j=1}^{23}\Gamma(m+{j\over24})z^m(-)^{m+1}\over\Gamma(m+{1\over2})
\prod_{k=-2}^5\Gamma(m+{k\over8})\prod_{l=-10}^1\Gamma(m+{l\over10})m^2m!(m-1)!}\nonumber\\
& & \times\biggl[\sum_{j=1}^{23}\Psi(m+{j\over24})+ln(-z)-1-\Psi({1\over2}+m)
-\sum_{k=-2}^5\Psi(m+{k\over8})-\sum_{l=10}^2\Psi({l\over12}+m)-{1\over m}\nonumber\\
& & -\sum_{j=1}^m{1\over j}-\Psi(m+1)
\biggr]
\end{eqnarray}

Similarly, for $|z|>1$, by deforming the contour $\gamma$ to the one $\gamma^{\prime\prime}$,
and on using that $Res(\Gamma({j\over24}+s)|_{s=-m(\in{\bf Z}^-\cup{0})-{j\over24}}
={(-)^m\over m!}$, one gets the following result:
\begin{equation}
\label{eq:sol1largez}
\sum_{m=0}^\infty\sum_{j=1}^{23}{-\Gamma({j\over24}+m)(-)^m\prod_{j\neq k=1}^{23}
\Gamma({k-j\over24}-m)\over m!(m+{j\over24})\Gamma(-m-{j\over24}+1)\prod_{k=-2}^8\Gamma
({k\over8}-m-{j\over24})\prod_{l=10}^1\Gamma({l\over12}-m-{j\over24})}(-z)^{-m-{j\over24}}.
\end{equation}
We explicitly evaluate below the solutions $(a_1),(a_3),(a_{12})$, the others following
suit.

($a_1$)
\begin{eqnarray}
\label{eq:sol2}
& & I\left(\begin{array}{c|c}\\
00& 0{1\over24}...{23\over24}\\
&\nonumber\\ \hline
&\nonumber\\ 
. & 1{1\over2}{5\over8}...-{2\over8}{1\over12}...-{10\over12}\\
\end{array}\right)(z)\nonumber\\
& & ={1\over2\pi i} \int_{\gamma^\prime}{[\Gamma(-s)]^2\Gamma(s)\prod_{j=1}^{23}\Gamma({j\over24}+s)\over
\Gamma(1+s)\Gamma({1\over2}+s)\prod_{k=-2}^5\Gamma({i\over8}+s)\prod_{l=10}^1\Gamma({l\over12}+s)}
(-z)^s\nonumber\\
& & 
\end{eqnarray}
For arguments similar to the one in (a) above, again the pole of order 3 at $s=0$ does not contribute
to the residue \footnote{For the sake of completeness, the finite term that gets multiplied by
${1\over\Gamma(0)^2}$ this time is:
$\biggl[\sum_{j=1}^{23}\Psi({j\over24})+ln(z)+2\gamma-2
-\sum_{k=-2}^5\Psi({k\over8})-\sum_{l=-{10}}^1\Psi({l\over12})\biggr]^2$ 
$+2\Psi^\prime(1)+\sum_{j=1}^{23}\Psi^\prime({j\over24})+4
-\sum_{k=-2}^5\Psi^\prime({k\over8})-\sum_{l=10}^1\Psi^\prime({l\over12})$.}.The final answer is:
\begin{eqnarray}
\label{eq:sol2fin}
& & \sum_{M=1}^\infty {1\over (m!)^2}{\prod_{j=1}^{23}\Gamma({j\over24}+m)\over\Gamma(m+{1\over2})
\prod_{k=-2}^5\Gamma({k\over8}+m)\prod_{l=10}^1\Gamma({l\over12}+m)}(z)^m\nonumber\\
& & \times\biggl[2\gamma+\sum_{j=1}^{23}\Psi({j\over24}+m)-\Psi(m+1{1\over2})-{1\over m}-
\sum_{k=-2}^5\Psi({k\over8}+m)-\sum_{l=1}^{10}\Psi({l\over12}+m)+ln(z)\nonumber\\
& & -2\sum_{j=1}^m{1\over j}\biggr]
\end{eqnarray}

Similar to $(a)$ above, 
for $|z|>1$, one gets the following result:
\begin{equation}
\label{eq:sol2largez}
\sum_{m=0}^\infty\sum_{j=1}^{23}{(\Gamma({j\over24}+m))^2(-)^m\prod_{j\neq k=1}^{23}
\Gamma({k-j\over24}-m)\over m!(m+{j\over24})\prod_{k=-2}^8\Gamma
({k\over8}-m-{j\over24})\prod_{l=10}^1\Gamma({l\over12}-m-{j\over24})}z^{-m-{j\over24}}.
\end{equation}

($a_3$)
\begin{eqnarray}
\label{eq:sol4}
& & I\left(\begin{array}{c|c}\\
000{1\over2} & 0{1\over24}...{23\over24}\\
&\nonumber\\ \hline
&\nonumber\\ 
.&{5\over8}...-{2\over8}{1\over12}...-{10\over12}\\
\end{array}\right)(z)\nonumber\\
& & ={1\over2\pi i}\int_{\gamma^\prime}{[\Gamma(-s)]^3\Gamma(s)\prod_{j=1}^{23}\Gamma({j\over24}+s)
\Gamma({1\over2}-
)\over \prod_{k=-2}^5\Gamma({i\over8}+s)\prod_{l=10}^1\Gamma({l\over12}+s)} (z)^s
\end{eqnarray}
Using $\Gamma(-s)\Gamma(s)={\pi\over s\ sin(\pi s)}$, and arguments above, one can show that the
pole of order 4 at $s=0$ does not contribute to the residue, and the pole of order 3 at $s=m\in{\bf Z}^+$
and a simple at $s=n(\in{\bf Z}^+\cup\{0\})+{1\over2}$,
contribute
 the following to the residue:
\begin{eqnarray}
\label{eq:sol4fin}
& & \sum_{m=1}^\infty{z^m\over(m!)^2}{\prod_{j=1}^{23}\Gamma({j\over24}+m)\over\prod_{k=-2}^5
\Gamma({k\over8}+m)\prod_{l=10}^1\Gamma({l\over12}+m)}\nonumber\\
& & \times\biggl[\biggl\{2\gamma-2\sum_{j=1}^m{1\over j}+\sum_{j=1}^{23}\Psi({j\over24}+m)-
\sum_{j=-2}^5\Psi({j\over8}+m)-\sum_{j=10}^1\Psi({j\over12}+m)+ln(z) \biggr\}^2
\nonumber\\
& & -2\Psi^\prime(1)+2\sum_{j=1}^m{1\over j^2}+{\pi^2\over3}
+\sum_{j=1}^{23}\Psi^\prime({j\over24}+m)-\sum_{j=-2}^5\Psi^\prime({j\over8}+m)-\sum_{j=-10}^1\Psi^\prime
({l\over12}+m)
\biggr]\nonumber\\
& & \sum_{m=0}^\infty{z^{m+{1\over2}}(-)^{m+1}\over m!}{[\Gamma(-m-{1\over2})]^3\Gamma(m+{1\over2})
\prod_{j=1}^{23}\Gamma(m+{j\over24}+{1\over2})\over\prod_{k=-2}^5\Gamma({k\over8}+{1\over2}+m)
\prod_{l=-10}^1\Gamma({l\over12}+{1\over2}+m)}
\end{eqnarray}

For $|z|>1$, one gets the following result:
\begin{equation}
\label{eq:sol3largez}
\sum_{m=0}^\infty\sum_{j=1}^{23}{-(\Gamma({j\over24}+m))^2\pi\Gamma({1\over2}+{j\over24}+m)
(-)^m\prod_{j\neq k=1}^{23}
\Gamma({k-j\over24}-m)\over ({j\over24}+m)sin[\pi({j\over24}+m)]m!(m+{j\over24})\prod_{k=-2}^8\Gamma
({k\over8}-m-{j\over24})\prod_{l=10}^1\Gamma({l\over12}-m-{j\over24})}z^{-m-{j\over24}}.
\end{equation}

($a_{12}$)
\begin{eqnarray}
\label{eq:sol12}
& & I\left(\begin{array}{c|c}\\
0 0 0 {1\over2} {3\over8} ....{10\over8} {11\over12} &
0 {1\over24}....{23\over24}\\
& \\ \hline
& \\
.&0 -{1\over12}...-{10\over12}\\
\end{array}\right)(-z)\nonumber\\
& & ={1\over2\pi i}\int_{\gamma^\prime}{[\Gamma(-s)]^3\Gamma({1\over2}-s)\prod_{n=3}^{10}
\Gamma({n\over8}-s)\Gamma({11\over12}-s)\prod_{j=1}^{13}\Gamma({j\over24}+
s)\over\prod_{j=1}^{10}\Gamma(-{k\over12}+s)}(-z)^s
\nonumber\\
& & 
\end{eqnarray}
The integral (\ref{eq:sol12}) has a pole of order 3 at $s=0$ and simple poles
at $s={1\over2}+m,{n\over8}+m,{11\over12}+m,m\in{\bf Z}^+\cup\{0\}$, and the final
answer is:
\begin{eqnarray}
\label{eq:sol12fin}
& & \sum_{m=0}^\infty{-(-)^m(-z)^m\over m!}{\Gamma({1\over2}-m)\prod_{n=3}^{10}
\Gamma({n\over8}-m)\Gamma({11\over12}-s)\over\prod_{k=1}^{10}\Gamma(-{k\over12}+m)}
\biggl[\{-3\gamma-3\sum_{j=1}^m{1\over j}-\Psi({1\over 2}-m)+ln(-z)\nonumber\\
& & -\sum_{n=3}^{10}\Psi({n\over8}-m)-\Psi({11\over12}-m)-\sum_{k=1}^{10}
\Psi(-{k\over12}+m)\}^2\nonumber\\
& & +3\Psi^\prime(1)+3\sum_{j=1}^m{1\over j^2}+\psi^\prime({1\over2}-m)
-\sum_{n=3}^{10}\Psi^\prime({n\over8}-m)-\Psi^\prime({11\over12}-m)-
\sum_{k=1}^{10}\Psi^\prime(-{k\over12}+m)\biggr]\nonumber\\
& & +\sum_{m=0}^\infty{(-)^{m+1}(-z)^{m+{1\over2}}\over m!}{\Gamma(-m-{1\over2})
\prod_{n=3}^{10}\Gamma({n\over8}-m-{1\over2})\Gamma({5\over12}-m)\over
\prod_{k=1}^{10}\Gamma(-{k\over12}+m+{1\over2})}\nonumber\\
& & +\sum_{m=0}^\infty\sum_{n=3}^{10}{(-)^{m+1}(-z)^{m+{n\over8}}\over m!}{[\Gamma(
-m
-{n\over8})]^3\Gamma({1\over2}-m-{n\over8})\prod_{n^\prime=3\neq n}^{10}\Gamma({n^\prime}
-{n\over8}-m)\Gamma({11\over12}-{n\over8}-m)\over\prod_{k=1}^{10}\Gamma(-{k\over12}+m
+{n\over8})}\nonumber\\
& & +\sum_{m=0}^\infty{(-)^{m+1}(-z)^{{11\over12}+m}\over m!}{[\Gamma(-{11\over12}-m)]^3
\Gamma(-{5\over6}-m)\prod_{n=3}^{10}\Gamma({n\over8}-{11\over12}-m)\over\prod_{k=1}^{10}
\Gamma(-{(k-11)\over12}+m)}
\end{eqnarray}

For $|z|>1$, one gets the following result:
\begin{equation}
\label{eq:sol12largez}
\sum_{m=0}^\infty\sum_{j=1}^{23}
{(\Gamma({j\over24}+m))^3\pi\Gamma({1\over2}+{j\over24}+m)
(-)^m\prod_{j\neq k=1}^{23}
\Gamma({k-j\over24}-m)\prod_{n=3}^{10}
\Gamma({n\over8}+{j\over24}+m)\Gamma({11\over12}+{j\over24}+m)
\over m!(m+{j\over24})
\prod_{l=10}^1\Gamma({l\over12}-m-{j\over24})}z^{-m-{j\over24}}.
\end{equation}
The appearance of log in the above solutions is indicative of the degeneracy
in the indicial equation corresponding to the Picard-Fuchs equation. Note
the absence of $(ln z)^3$ terms in the solutions above.

The Picard-Fuchs equation can be written in the form:
\begin{equation}
\label{eq:PFform}
\biggl(\Delta_z^{24}+\sum_{i=1}^{23}{\bf B}_i(z)\Delta_z^i\biggr)\tilde{\Pi}(z)
=0,
\end{equation}
where the 23 ${\bf B}_i$'s after rescaling $z$ such that the coefficient of
$\Delta_z^{24}$ is $1-z$, is given in Appendix A.
The Picard-Fuchs
equation in the form written in (\ref{eq:PFdiffeq1}) can alternatively be expressed as
the following system of 24 linear differential equations:
\begin{eqnarray}
\label{eq:PFdiffeq3}
& & \Delta_z\left(\begin{array}{c}\\
\tilde{\Pi}(z)\\
\Delta_z 
\tilde{\Pi}(z)\\
(\Delta_z)^2 
\tilde{\Pi}(z)\\
...\\
(\Delta_z)^{23} 
\tilde{\Pi}(z)\\
\end{array}\right)
=\nonumber\\
& & \left(\begin{array}{ccccc}\\
0 & 1 & 0 & ... 0 & 0\\
0 & 0 & 1 & ... 0 & 0\\
. & . & . & ... . & . \\
0 & 0 & 0 & ... 0 & 1 \\
0 & -{\bf B}_1(z) & -{\bf B}_2(z) & ... -{\bf B}_{22}(z) & -{\bf B}_{23}(z)\\
\end{array}\right)
\left(\begin{array}{c}\\
\tilde{\Pi}(z)\\
\Delta_z 
\tilde{\Pi}(z)\\
(\Delta_z)^2 
\tilde{\Pi}(z)\\
...\\
(\Delta_z)^{23} 
\tilde{\Pi}(z)\\
\end{array}\right)
\end{eqnarray}
The matrix on the RHS of (\ref{eq:PFdiffeq3}) is usually denoted by $A(z)$. 

If the 24 solutions, $\{\tilde{\Pi}_{I=1,...,24}\}$,
 are collected as a column vector $\tilde{\Pi}(z)$, then the
monodromy matrix $T$ for $|z|<<1$ is defined by:
\begin{equation}
\tilde{\Pi}(e^{2\pi i}z)=T\tilde{\Pi}(z).
\end{equation}
The basis for the space of solutions can be collected as the
columns of the ``fundamental matrix" $\Phi(z)$ given by:
\begin{equation}
\label{eq:PFdiffeqsol}
\Phi(z)=S_{24}(z)z^{R_{24}},
\end{equation}
where $S_{24}(z)$ and $R_{24}$ are 24$\times$24 matrices that single and
multiple-valued respectively. Note that ${\bf B}_i(0)\neq0$, which influences
the monodromy properties. Also,
\begin{equation}
\label{eq:Phidieuf2}
\Phi(z)_{ij}=\left(\begin{array}{ccc}\\
\tilde{\Pi}_1(z) & ... & \tilde{\Pi}_{24}(z) \\
\Delta_z \tilde{\Pi}_1(z) & ... & \Delta \tilde{\Pi}_{24}(z) \\
\Delta_z^2 \tilde{\Pi}_2(z) & ... & \Delta^2 \tilde{\Pi}_{24}(z) \\
... & ... & ... \\
\Delta_z^{23} \tilde{\Pi}_1(z) & ... & \Delta_z^{23} \tilde{\Pi}_{24}(z)\\
\end{array}\right)_{ij},
\end{equation} 
implying that
\begin{equation}
\label{eq:Tdieuf}
T=e^{2\pi i R^t}.
\end{equation}
Now, writing $z^R=e^{R ln z}=1 + R ln z + R^2(ln z)^2+...$, and further
noting that there are no terms of order higher than $(ln z)^2$ in 
$\tilde{\Pi}(z)$ 
obtained above, implies that the matrix $R$ must satisfy the property:
$R^{m}=0,\ m=3,4,...\infty$. Hence, $T=e^{2\pi i R^t}=1 + 2\pi i R^t +
{(2\pi i)^2\over2}(R^t)^2.$ It is not possible to evaluate all the eigenvalues
of the matrix $A(0)$ (using the expressions for $B_i(z)$ in Appendix A) using
Mathematica because of the degree-24 characteristic
equation that one would require to solve. Mathematica does predict a 4-fold
degenerare null eigenvalue though. Irrespective of whether or not the
distinct eigenvalues of $A(0)$ differ by integers, one has to evaluate
$e^{2\pi i A(0)}$. Using Mathematica, one gets a very complicated expression,
whose form in terms of powers of $\pi$ and the null entries are given by:
\begin{eqnarray}
\label{eq:expA(0)1}
& & e^{2i\pi A(0)}=\nonumber\\
& & \left(\begin{array}{ccccccccc}\\
1&2i\pi&(..)\pi^2&(..)\pi^3&(..)\pi^4&(..)\pi^5&(..)\pi^6&...(..)\pi^{22}&(..)
\pi^{23}\\
0&1&2i\pi&-2i\pi^2&-{4i\pi^3\over3}+
&(..)\pi^4+&(..)\pi^5+&...(..)\pi^{21}+&(..)\pi^{22}\\
& & & & (..)\pi^{23}&(..)\pi^{23}&(..)\pi^{23}&(..)\pi^{23}
&\\
0&0&1&2i\pi&-2\pi^2
&(..)\pi^3&(..)\pi^4+&(..)\pi^{20}+&(..)\pi^{21}+\\
&&&&+(..)\pi^{22}&+(..)\pi^{22}&(..)\pi^{22}+&(..)\pi^{22}+&(..)\pi^{23}\\
&&&&&+(..)\pi^{23}&(..)\pi^{23}&(..)\pi^{23}&\\
0&0&0&1&2i\pi&(..)\pi^2+&(..)\pi^3+&...(..)\pi^{19}&(..)\pi^{20}+\\
&&&&&(..)_i\pi^{21}&\sum_{i=21}^{23}(..)_i\pi^i
&+\sum_{i=21}^{23}(..)_i\pi^i&(..)\pi^{22}\\
&&&&&&&&+(..)\pi^{23}\\
0&0&0&0&1&2i\pi&(..)\pi^2&...(..)\pi^{18}&(..)\pi^{19}+\\
&&&&&&+(..)\pi^{20}&+\sum_{i=20}^{23}(..)_i\pi^i
&\sum_{i=21}^{23}(..)\pi^i\\
&&&&&&+(..)\pi^{22}&&\\
&&&&&&+(..)\pi^{23}&&\\
0&0&0&0&0&1&2i\pi+&...(..)\pi^{17}+&(..)\pi^{18}+\\
&&&&&+(..)\pi^{19}&\sum_{i=19}^{23}(..)_i\pi^i&
\sum_{i=19}^{23}(..)_i\pi^i
&\sum_{i=20}^{23}(..)_i\pi^i\\
..&..&..&..&..&..&..&..&..\\
0&0&0&0&0&(..)\pi^2+&....&...1+&2i\pi+\\
&&&&&\sum_{i=4}^{23}(..)_i\pi^i&....&
\sum_{i=2}^{23}(..)_i\pi^i&
\sum_{i=3}^{23}(..)_i\pi^i\\
0&0&0&0&0&e^{i\alpha_1}((..)
\pi+&e^{i\alpha_2}\times&...e^{i\alpha_{17}}\times&-1+\\
&&&&&
\sum_{i=3}^{23}(..)_i\pi^i)&\sum_{i=1}^{23}(..)_i\pi^i
&\sum_{i=1}^{23}(..)_i\pi^i&+\sum_{i=2}^{23}(..)_i\pi^i\\
\end{array}\right)\nonumber\\
& & 
\end{eqnarray}
Under the change of basis (See \cite{GL}): 
$U(z)\rightarrow U^\prime(z)=M^{-1}U(z)$, and by choosing $M$ such that
$S^\prime(0)={\bf 1}_{24}$, one gets $T=Me^{2i\pi A(0)}M^{-1}$. In principle,
one could try to evaluate $M$ using techniques given in \cite{GL}.

\subsubsection{The MATHEMATICAlly interesting point $e^t=2229025112064$}

We now discuss Picard-Fuchs equation about a certain point, where curiously,
using Mathematica, one is able to solve for the eigenvalues of the 
24$\times$24 matrix ``$A(0)$'', and one finds that they differ by integers
implying that the monodromy matrix can not be given by $e^{2i\pi A(0)}$.
There is not much known about solutions to the Picard-Fuchs equation 
about $z(\equiv e^{-t^\prime})=1$, and after rescaling, the abovementioned
point, in terms of the rescaled $z$, corresponds precisely to solving the
Picard-Fuchs equation about (rescaled)$z=1$. To be able say something definite
about the monodromy, will help in understanding the solution as well. It
is in this regard that this (sub)subsection is quite relevant.
 
If one does not make the substitution $e^{t^\prime}=z$, then
the Picard-Fuchs equation can be written in the following form:
\begin{equation}
\label{eq:PFdiffeq0}
\biggl[(e^{t^\prime} - 2229025112064){d^{24}\over dt^\prime\ ^{24}}
+\sum_{i=1}^{23}\tilde{C}_i(t^\prime){d^i\over dt^\prime\ ^i}
\biggr]\tilde{\Pi}(t^\prime)=0,
\end{equation}
where $\tilde{C}_i(t^\prime)$'s are regular. 
After the shift: $t^\prime\rightarrow t^\prime+ln(2229025112064)$, and using:
\begin{equation}
\label{eq:lnder}
t^\prime\ ^m{d^m\over dt^\prime\ ^m}=\prod_{n=0}^{m-1}(t^\prime{d\over dt^\prime}-n),
\end{equation}
one can rewrite the Picard-Fuchs equation (\ref{eq:PFdiffeq0}) as:
\begin{equation}
\label{eq:PFdiffeq1}
\biggl[\biggl(t^\prime{d\over dt^\prime}\biggr)^{24}+\sum_{i=1}^{23}{\bf B}_i(t^\prime)\biggl(t^\prime{d\over dt^\prime}\biggr)^i\biggr]
\tilde{\Pi}(t^\prime)=0,
\end{equation}
where
\begin{equation}
\label{PFdiffeq2}
{\tilde{C}_i(t^\prime)\over{22902511064(e^{t^\prime}-1)}}\equiv 
{\bf B}_i(t^\prime) = b^0_i+{\sum_{j=1}^{24-i}(b^1_j
+e^{t^\prime}b^2_j)t^\prime\ ^j\over(e^{t^\prime}-1)}.
\end{equation}
One hence notices that $t^\prime=0$ is a regular singular point of the differential equation
(\ref{eq:PFdiffeq1}).
The expressions for the 24 ${\bf B}_i's$ (from where one can read off $b^{0,1,2}_j$'s) is
given partly for the sake of completeness and partly to verify the structure of terms
as given on the RHS of (\ref{eq:PFdiffeq1}) and to read off the expressions for ${\bf B}_i(0)'s$
necessary for determining the matrix $A(0)$, is given in the appendix. 
The matrix $A(0)$ is given by:
\begin{eqnarray}
\label{eq:A(0)}
& & A(0)=\nonumber\\
& & \left(\begin{array}{cccc}\\
0 & 1 & 0 &  ... 0 \\
0 & 0 & 1 & ... 0 \\ 
. & . & . & ...   \\
2248001455555215360000 & -9420954286448517120000 & 17660770126877316096000 & ... 255 \\
\end{array}\right)\nonumber\\
& & 
\end{eqnarray}
The reason why $t^\prime$(before the shift)=$ln(2229025112064)$ is MATHEMATICAlly interesting is
that Mathematica is able to determine the eigenvalues of $A(0)$ ! They are given by
$0,1,(2)^2,3,...,23$ - they differ by integers. As a consequence, one can not
set $R_{24}$ to $A(0)$ (See \cite{Morrison}). 

The matrix $e^{2i\pi A(0)}$ is given by:
\begin{equation}
\label{eq:expA(0)z=1}
e^{2i\pi A(0}=\left(\begin{array}{ccccc}\\
1&i\pi X_1&-i\pi X_2 & ...& i\pi X_{23}\\
0&1+2i\pi X_1 & -2i\pi X_2 & ... & 2i\pi X_{23} \\
0&2^2i\pi X_1 & 1 - 2^2i\pi X_2 & ...& 2^2i\pi X_{23} \\
...&...&...&...&...\\
0&2^{23}i\pi X_1&2^{23}i\pi X_2&...&1+2^{23}i\pi X_{23}\\
\end{array}\right)
\end{equation}
where

$\{X_1,\ X_2,\ X_3,\ X_4,\ X_5,\ X_6,\ X_7,\ X_8,\ X_9,\ X_{10},\
X_{11},\ X_{12},\ X_{13},\ X_{14},\ X_{15},\ X_{16},$
\vskip 0.1in
$ X_{17},\ X_{18},\ 
X_{19},\ X_{20},\ X_{21},\ X_{22},\ X_{23}\}$
\vskip 0.in
$=
\{ 462\,i \,\pi ,\frac{-57279591\,i }{33592}\,\pi ,\frac{7757553775549\,i }{2793510720}\,\pi ,\frac{-287102134746031\,i }{106661318400}\,\pi ,
  \frac{3311858525015123\,i }{1882258560000}\,\pi ,\frac{-2399093587319011\,i }{2887073280000}\,\pi ,$
\vskip 0.1in
$\frac{1282161754601213\,i }{4330609920000}\,\pi,
  \frac{-100177907111227\,i }{1222760448000}\,\pi ,\frac{54926206286431\,i }{3056901120000}\,\pi ,\frac{-542316429868939\,i }{171186462720000}\,\pi ,
  \frac{38813357767589\,i }{85593231360000}\,\pi ,\frac{-26600161411\,i }{501645312000}\,\pi ,$
\vskip 0.1in
$\frac{140203969733\,i }{27590492160000}\,\pi ,
  \frac{-39051524303\,i }{97820835840000}\,\pi ,\frac{2508239171\,i }{97820835840000}\,\pi ,\frac{-91643431\,i }{68474585088000}\,\pi ,
  \frac{19239707\,i }{342372925440000}\,\pi ,\frac{-517891\,i }{277159034880000}\,\pi ,$
\vskip 0.1in
$\frac{4439\,i }{92386344960000}\,\pi ,
  \frac{-5819\,i }{6319225995264000}\,\pi ,\frac{23\,i }{1858595880960000}\,\pi ,\frac{-23\,i }{221172909834240000}\,\pi ,
  \frac{i }{2432902008176640000}\,\pi \}.$
\vskip 0.1in
Again, as in {\bf 3.1.2}, for a suitable ${\cal M}$, the monodromy matrix
is given by $T={\cal M}e^{2i\pi A(0)}{\cal M}^{-1}$.
\section{Superpotential Calculation}

In \cite{tri1}, at the level of spectrum matching, it was shown that Heterotic on the self-mirror 
Calabi-Yau $CY_3(11,11)$ is dual to $M$ theory on the `barely' $G_2$-Manifold ${CY_3(3,243)\times S^1\over{\bf Z}_2}$
where the ${\bf Z}_2$ involved the Enriques involution times a reflection of the $S^1$. It was also shown
that the expected $F$-theory dual for the above must involve an elliptically fibered $CY_4$ that 
has the Hodge data: $h^{1,1}=12, h^{2,1}=128,h^{3,1}=108$ and which has a trivially rationally ruled
base given by ${\bf CP}^1\times$Enriques surface\footnote{As mentioned in
the footnote on page 2, there was a puzzle raised in the context of
the $F$-theory dual, namely, the required 4-fold could neither be obtained as a free involution (to guarantee
a null Euler characteristic, because there are no $F$-theory 3-branes as there are no nonperturbative
Heterotic 5-branes in Heterotic on $CY_3(11,11)$) of $CY_3(3,243)\times T^2$, nor is it possible to 
construct it as a hypersurface in a toric variety explaining its absence in the (incomplete) list of
$CY_4$'s obtained as hypersurfaces in ${\bf WCP}^5$ in \cite{Kreuz}.}.

In this section, we first give a heuristic argument to show that there can no be no superpotential generated on
the type IIA side that got uplifted the abovementioned `barely'
$G_2$-manifold, by indicating that there
can be no superpotential on the $F$ theory and Heterotic sides. Then, using mirror symmetry, away from
the orbifold singularities of $CY_3(3,243)$, we show explicitly that indeed, there is no superpotential
on the freely acting orbifold of $CY_3(3,243)$ as used in \cite{VW,tri1}.

As per the work of Witten in \cite{nonpertW}, if one considers $F$-theory on an
elliptically fibered $CY_4$ $X_4$ with the holomorphic map $\pi:X_4\rightarrow B_3$ having a 6-divisor $D_3$
as a section such that $\pi(D_3)=C_2\subset B_3$, then in the limit of vanishing size of the the elliptic
fiber, 5-branes wrapped around $D_3$ in $M$-theory on the same $X_4$ obeying the unit-arithmetic genus
condition:
\begin{equation}
\label{eq:chi=1}
\chi(D_3,{\cal O}_{D_3})=1,
\end{equation}
correspond to 3-branes wrapped around $C_2$ in type $IIB$, or equivalently $F$-theory 3-branes wrapped around
$C_2\subset B_3$. In other words, only 3-branes contribute to the superpotential in $F$-theory. Since,
there are no 3-branes in the $F$-theory dual that we have, this implies that there can be no superpotential
generated on the $F$-theory side.

Now, given that F-theory 3-branes correspond to Heterotic instantons, one again expects no superpotential to
be generated on the Heterotic theory side on the self-mirror $CY_3(11,11)$ based on the ${\cal N}=2$ type IIA/Heterotic
dual of \cite{Feretal} 
where the same self-mirror Calabi-Yau figured on the type IIA side and the self-mirror nature was argued to show
that there are no world-sheet or space-time instanton corrections to the classical moduli space.\footnote{We thank W.Lerche for pointing this out to us.} Now, if the 
abovementioned triality is correct, then one should be able to show that there is no superpotential generated
on type $IIA$ side on the freely-acting antiholomorphic involution of $CY_3(3,243)$. We will, in this paper,
look at the Fermat's hypersurface of section {\bf 2} away from the orbifold singularities discussed in the same
section. The freely acting antiholomorphic involution $\omega$ is defined as:
\begin{equation}
\label{eq:invdieuf}
\omega:(z_1,z_2,z_3,z_4,z_5)\rightarrow({\bar z}_2,-{\bar z}_1,{\bar z}_3,{\bar z}_4,{\bar z}_5).
\end{equation}
As discussed in \cite{tri1}, $\omega$ has the interesting property that as an action on the cohomology of
$CY_3(3,243)$, it reflects $H^{1,1}$ and complex conjugates $H^{2,1}$ in general, and in the large 
base ($\equiv{\bf CP}^1$) limit of $CY_3(3,243)$, as in \cite{VW}, $\omega$ reflects the K\"{a}hler form
and complex conjugates every element of $H^{2,1}$ {\it precisely}, i.e., not just up to total differentials.

In the following,
the arguments similar to those of \cite{AAHV} will used.
Now, on the mirror type $IIB$ side, the superpotential $W$ is generated from domain-wall tention, the domain wall
being $D5$-branes wrapped around supersymmetric 3-cycles embedded in Calabi-Yau 3-folds. The expression for the
superpotential is given by:
\begin{equation}
W_{IIB}=\int_{C:\partial C=\sum_i D_i}\Omega_3,
\end{equation}
where $\Omega_3$ is the holomorphic 3-form for the Calabi-Yau 3-fold, and
$D_i$'s are 2-cycles corresponding to the positions of $D5$-branes or $O5$-planes, i.e., objects carrying
$D5$ brane charge. From the world-sheet point of view, the $D5$ branes correspond to disc amplitudes and
$O5$-planes correspond to ${\bf RP}^2$ amplitudes\footnote{If $z_{1,2}$ are used for the homogenous coordinates of
the ${\bf CP}^1$ in the $CY_3(3,243)$ regarded as $K3$-fibration over ${\bf CP}^1$, then one gets an ${\bf RP^2}$
from ${{\bf CP}^1\over\omega}$.} As there are no branes in our theory, we need to consider only ${\bf RP}^2$
amplitudes. Now, type $IIA$ on a freely acting involution of a Calabi-Yau with no branes or fluxes can still
generate a superpotential because it is possible that free involution on type $IIA$ side corresponds to
orientifold planes in the mirror type $IIB$ side, which can generate a superpotential. 
An example of this is that a freely acting orientifold of $IIB$ on $S^1$  involving a shift along the $S^1$ alongwith
world sheet orientation reversal
is T-dual to an $O8^+/O8^-$ pair  at antipodal points on the dual circle, 
as given  in \cite{9712028}\footnote{We thank M.Aganagic for pointing this out to us.}.

Now, the ${\cal N}=1$ Landau-Ginsburg superpotential $W_{LG}$ enters as:
\begin{equation}
\label{eq:WLG}
\int d^2\theta W_{LG}.
\end{equation}
Now, the measure is reflected under the world-sheet orientation reversal $\Omega$. Hence, for (\ref{eq:WLG}) to be
invariant under $\Omega.\omega$, 
\begin{equation}
\label{eq:Ww}
\omega:W_{LG}\rightarrow -W_{LG}.
\end{equation}
From the discussion in section {\bf 2}, one knows that 
\begin{equation}
W_{LG}=\sum_{i=1}^6e^{-Y_i},
\end{equation}
where $Y_i$'s are the twisted chiral superfields (of the mirror to the $U(1)$ gauged linear sigma model). Now,
promoting the action of $\omega$ given in (\ref{eq:invdieuf}) to the one on the chiral superfields $X_{i=1,...,6}$:
\begin{equation}
\label{eq:omegadieuf}
\omega:(X_1,X_2,X_3,X_4,X_5,X_6)\rightarrow({\bar X}_2,-{\bar X}_1,{\bar X}_3,{\bar X}_4,{\bar X}_5,{\bar X}_6),
\end{equation}
and using $Re(Y_i)=|X_i|^2$, one gets the following action of $\omega$ on the twisted chiral superfields $Y_i$'s:
\begin{eqnarray}
\label{eq:omegaY}
& & \omega:Y_1\rightarrow Y_2+i\pi,\ Y_2\rightarrow Y_1+i\pi;\nonumber\\
& & Y_{3,4,5,6}\rightarrow Y_{3,4,5,6}+i\pi.
\end{eqnarray}
The action of $\omega$ on $Y_{3,4,5,6}$ implies that $\omega$ acts without fixed points even on the twisted
chiral superfields, further implying that there are no orientifold fixed planes, and {\it hence no
superpotential is generated on the type $IIA$ side away from the orbifold
singularities}.

\section{Conclusion and Future Directions}

In this paper, we considered type $IIA$ and a free-orientifold of it on 
the compact $CY_3(3,243)$ expressed as a degree-24 hypersurface
in ${\bf WCP}^4[1,1,2,8,12]$, away from its orbifold singularities, from a 
gauged linear sigma model
point of view. We first derived the Picard-Fuchs equation, and obtained its Meijer basis
of solutions in the large {\it and} small complex structure limits (on the 
mirror Landau-Ginsburg side) of the Calabi-Yau. We made some
comments about the monodromy properties of the solutions, including at the special point,
$|z(\equiv$ rescaled$|=1(\leftrightarrow$ ``original" $z=1/(2229025112064)$). Then, we argue both indirectly and then directly,
that there can be no superpotential generated in type $IIA$ on the aforementined free involution
of the Fermat hypersurface (that gets smoothed out to $CY_3(3,243)$). 
So far, the argument has been given in the moduli space of the Calabi-Yau
away from the  orbifold singularities of the hypersurface in ${\bf WCP}^4[1,1,2,8,12]$. 
One has to see, how to improve the argument to consider the resolved manifold with
a $K3$ fibration over a ${\bf CP}^1$, and the $K3$ itself being an elliptic fibration over another
${\bf CP}^1$. In this direction, one notes that after resolving the ${\bf Z}_2$ and
${\bf Z}_4$ singularities discussed in section ${\bf 2}$, one ends up introducing two
new chiral superfields corresponding to the two ${\bf CP}^1$'s
that are required to
be introduced in blowing up the singularities. One then has to consider three instead
of a single $C^*$ action, and the $CY_3(3,243)$
\footnote{The $CY_3(3,243)$ considered in this paper will be
 an elliptic fibration over the Hirzebruch surface $F_2$.}
can be expressed as a suitable 
holomorphic quotient corresponding to a smooth toric variety. To be more specific, one
considers the resolved Calabi-Yau $CY_3(3,243)$ as the holomorphic quotient:
${C^7-F\over(C^*)^3}|_{\rm
hyp\ constraint}$, where the diagonal $(C^*)^3$ actions
on the seven coordinates of $C^7$ are given by:
\begin{equation}
\label{eq:3Q's}
x^j\sim \lambda^{iQ^a_j}x^j,\ {\rm no\ sum\ over}\ j;\ a=1,2,3,
\end{equation}
where the three sets of charges $\{Q^{a=1,2,3}_{i=(0,),1,...,7}\}$ (the "0" being for
the extra chiral superfield with $Q^0_i=-\sum_{i=1}^7Q^a_i$) are give by the following
\footnote{We are deeply grateful to A.Klemm for discussions on material presented here 
on the resolution.}:
\begin{equation}
\label{eq:3C*'s}
\begin{array}{ccccccccc}\\ 
& {\cal X}_0 &{\cal X}_1 & {\cal  X}_2 & {\cal  X}_3 & {\cal X}_4 & {\cal X}_5 
& {\cal X}_6 & {\cal X}_7\\ \hline
Q^{(1)}_i:& 0 & 1 & 1 & -2
& 0 & 0 & 0 & 0 \\
Q^{(2)}_i:& 0 & 0 & 0 & 1 & 1 & 0 & 0 & -2 \\
Q^{(3)}_I: & -6 & 0 & 0 & 0 & 0 & 2 & 3 & 1 \\
\end{array}
\end{equation}
where on noting:
\begin{equation}
\label{eq:linrel}
Q^{(1)}+2Q^{(2)}+4Q^{(3)}=\begin{array}{cccccccc}
-24&1&1&0&2&8&12&0
\end{array},
\end{equation}
one identifies ${\cal X}_{3,7}$ as the two extra chiral superfields introduced as a consequence of
singularity resolution. 

To write the superpotential, one can start with the ansatz:
\begin{equation}
\label{eq:Wanst}
W=\sum_{e_{0,1,...,7}}a_{e_{0,1,...,7}}\prod_{i=0}^7{\cal X}^{e_i}_i,
\end{equation}
where in order to guarantee invariance under each of the three $C^*$ actions, one gets
the following three constraints on the eight chiral superfields:
\begin{eqnarray}
\label{eq:const}
& & e_1+e_2-2e_3=0\nonumber\\
& & e_3+e_4-2e_7=0\nonumber\\
& & -6e_0+2e_5+3e_6+e_7=0,
\end{eqnarray}
in which by keeping, say, $e_{0,1,2,5,6}$ arbitrary, one gets:
\begin{equation}
\label{eq:W}
W=\sum_{e_{0,1,2,5,6}}a_{e_{0,1,,2,5,6}}{\cal X}_0^{e_0}{\cal X}_1^{e_1}{\cal X}_2^{e_2}
{\cal X}_3^{{e_1+e_2\over2}}{\cal X}_4^{-{e_1+e_2\over2}+2[6e_0-(2e_5+3e_6)]}
{\cal X}_5^{e_5}{\cal X}_6^{e_6}{\cal X}_7^{6e_0-(2e_5+3e_6)},
\end{equation}
where one has to ensure that the resulting $W$ is transverse. One could perhaps, set 
$e_0=1$, so as to rewrite $W={\cal X}_0\tilde{W}$, as in \cite{Witten}.
The form of $W$ will help determine the antiholomorphic involution of the smooth $CY_3(3,243)$
for the ${\cal N}=1$ type IIA theory.

On the mirror Landau-Ginsburg side, for the ${\cal N}=2$ type IIB theory,
one notes that it will be given by a $U(1)^3$ gauge
theory with field strengths $F^{(a=1,2,3)}$ with a superpotential $\tilde{W}_{LG}$ given by:
\begin{equation}
\label{eq:WLG3}
\tilde{W}_{LG}=\sum_{a=1}^3F^{(a)}(\sum_{i=0}^7Q^{(a)}_iY_i-t^a)+(W_{LG}\equiv)
\sum_{i=0}^7e^{-Y_i}
\end{equation}
which after integrating out the $F$'s, gives the three constraints:
\begin{eqnarray}
\label{eq:const2}
& & Y_1+Y_2-2Y_3=t_1\nonumber\\
& & Y_3+Y_4-2Y_7=t_2\nonumber\\
& & -6Y_0+2Y_5+3Y_6+Y_7=t_3.
\end{eqnarray}
One thus gets:
\begin{equation}
W_{LG}=\sum_{i=0,1,2,5,6}e^{-Y_i}+e^{-{t_1\over2}}e^{-{Y_1+Y_2\over2}}
+e^{-{t_1\over2}-t_2}e^{{Y_1+Y_2\over2}-2[6Y_0-2Y_5-3Y_6]}+e^{-t_3}e^{-[6Y_0-2Y_5-3Y_6]}.
\end{equation}

One can also look at the special Lagrangian submanifolds on the type IIA
side and their mirrors on the type IIB side. We will do so in the context
of ${\cal N}=2$ theories.
The special Lagrangian submanifolds (See \cite{AV})
are characterized by charges ``$q_\alpha$''
which satisfy:
\begin{equation}
\label{eq:SLAG1}
\sum_i q^\alpha_i=0,
\end{equation}
and which restrict the chiral superfields ${\cal X}_i$'s to:
\begin{equation}
\label{eq:SLAG2}
\sum_i|{\cal X}_i|^2q_i^\alpha=c^\alpha,
\end{equation}
for some constant $c^\alpha$'s. Using $Re(Y_i)=|{\cal X}_i|^2$, one sees
the following equation for the mirror of the special Lagrangian 
submanifolds:
\begin{equation}
\label{eq:SLAG3}
\prod_i e^{-Y_i q^\alpha_i}=e^{-i\sum_i q^\alpha_i Im(Y_i)}e^{-c^\alpha}
\equiv\epsilon^\alpha e^{-c^\alpha}.
\end{equation}
Consider, for the resolved $CY_3(3,243)$,
\begin{equation}
\label{SLAG4}
q^1=(-1,1,0,0,0,0,0,0),
\end{equation}
implying the following special Lagragian submanifold:
\begin{equation}
\label{eq:SLAG5}
|{\cal X}_1|^2-|{\cal X}_0|^2=c^1.
\end{equation}
Now, using the following $D$-term constraint (there are three, of which one
considers the one involving ${\cal X}_0$):
\begin{equation}
\label{eq:Dconst1}
-6|{\cal X}_0|^2+2|{\cal X}_5|^2+3|{\cal X}_6|^2+|{\cal X}_7|^2=r^3,
\end{equation}
one gets the following Lagrangian submanifold equation:
\begin{equation}
\label{eq:SLAG6}
-6|{\cal X}_0|^2+2|{\cal X}_5|^2+3|{\cal X}_6|^2+|{\cal X}_7|^2=r^3-6c^1
\equiv\tilde{c}^1.
\end{equation}
Hence, from (\ref{eq:SLAG6}), one sees that an equivalent charge vector
is:
\begin{equation}
\label{eq:equivq1}
\tilde{q}^1=(0,-6,0,0,0,2,3,1).
\end{equation}
The mirror corresponding to the Lagrangian submanifold (\ref{eq:SLAG6}) and
(\ref{eq:equivq1}) is:
\begin{equation}
\label{eq:mirror}
e^{6Y_1-2Y_5-3Y_6+Y_7}=\epsilon^1e^{-\tilde{c}^1}.
\end{equation}

Returning back to the question of superpotential for the the ${\cal N}=1$
theory, what needs to be worked out is the antiholomorphic involution on the type IIA side, its
image on the mirror type IIB side, and one should then be able to show that just like
there were no O-planes and hence no superpotential 
away from the orbifold singularities before the singularity resolution,
shown in section {\bf 4}, there can be no superpotential generated even after resolving
the orbifold singularities.
Additionally, it will be interesting to evaluate and match the K\"{a}hler
potential on the Heterotic and $M$ theory sides, and to take its $F$-theory limit.

\section*{Acknowledgements}

We would like to thank  M.Aganagic and D-E.Diaconescu for useful communications
and clarifications, W.Lerche for discussion
and C.Lazaroiu for bringing \cite{GL} to our attention. Its a pleasure
to thank A.Klemm for very useful discussions. We also thank D.L\"{u}st for going through a 
preliminary version of the manuscript. This research work was
supported by the Alexander von Humboldt Foundation.

\appendix
\section{${\bf B}_i$'s for PF around $|z|<1$}

In this appendix, we give the expressions, obtained from Mathematica,
for the 23 ${\bf B}_i$'s that figure in the Picard-Fuchs equation
for points near $z=0$.

\begin{eqnarray}
\label{eq:Bist=0}
& & 
{\bf B}_1(z)=\frac{2505147019375\,z}{5385144351531158470656\,\left( -1 + z \right) } 
\nonumber\\
& & {\bf B}_2(z)= 
  \frac{149679187975625\,z}{3590096234354105647104\,\left( -1 + z \right) }\nonumber\\
& & {\bf B}_3(z)= 
  \frac{160288719172038625\,z}{96932598327560852471808\,\left( -1 + z \right) }\nonumber\\
& & {\bf B}_4(z)=  
  \frac{1897375316911910575\,z}{48466299163780426235904\,\left( -1 + z \right) }\nonumber\\
& & {\bf B}_5(z)= 
  \frac{\,\left( 3638250 + \frac{15191740560751946323\,z}{429981696} \right) }
   {56358560858112\,\left( -1 + z \right) }\nonumber\\
& &  {\bf B}_6(z)=\frac{\,\left( -214589025 + \frac{88158910649831337151\,z}{214990848} \right) }
   {56358560858112\,\left( -1 + z \right) }\nonumber\\
& & {\bf B}_7(z)=  
  \frac{\,\left( 5736960255 + \frac{96816543515895571753\,z}{26873856} \right) }
   {56358560858112\,\left( -1 + z \right) }\nonumber\\
& & {\bf B}_8(z)=
  \frac{\,\left( -92797915786 + \frac{331544607220340363311\,z}{13436928} \right) }
   {56358560858112\,\left( -1 + z \right) }\nonumber\\
& & {\bf B}_9(z)= 
  \frac{\,\left( 1021883387656 + \frac{2788771435559577853\,z}{20736} \right) }
   {56358560858112\,\left( -1 + z \right) }\nonumber\\
& & {\bf B}_{10}(z)=
  \frac{\,\left( -8165125067584 + \frac{55246113744858429349\,z}{93312} \right) }
   {56358560858112\,\left( -1 + z \right) }\nonumber\\  
& & {\bf B}_{11}(z)= \frac{\,\left( 49226099800176 + \frac{2756809280714517083\,z}{1296} \right) }
   {56358560858112\,\left( -1 + z \right) }\nonumber\\
& & {\bf B}_{12}(z)=  
  \frac{\,\left( -229685725831200 + \frac{4070384057007569521\,z}{648} \right) }
   {56358560858112\,\left( -1 + z \right) } \nonumber\\
& & {\bf B}_{13}(z)= 
  \frac{\,\left( 948851467387776 + 17223196429221419\,z \right) }
   {63403380965376\,\left( -1 + z \right) }\nonumber\\
& &  {\bf B}_{14}(z)= 
  \frac{\,\left( -2463325189511168 + \frac{277614451589633648\,z}{9} \right) }
   {56358560858112\,\left( -1 + z \right) }\nonumber\\
& & {\bf B}_{15}(z) = 
\frac{11\,\left( 2043640448992+ 18227621504883\,z \right) }{220150628352\,\left( -1 + z \right) }  \nonumber\\
& & {\bf B}_{16}(z)=  
  \frac{\,\left( -21034217831392 + 137272511800831\,z \right) }{110075314176\,\left( -1 + z \right) }  \nonumber\\
& & {\bf B}_{17}(z)=  
  \frac{\,\left( 13645602646 + 66672816001\,z \right) }{47775744\,\left( -1 + z \right) } 
\nonumber\\
& & {\bf B}_{18}(z)=  
  \frac{19\,\left( -425415838 + 1587345023\,z \right) }{23887872\,\left( -1 + z \right) } 
\nonumber\\
& & {\bf B}_{19}(z)=  
  \frac{361\,\left( 576063 + 1670053\,z \right) }{663552\,\left( -1 + z \right) }\nonumber\\
& & {\bf B}_{20}(z)= 
  \frac{\,\left( -73644729 + 168423871\,z \right) }{331776\,\left( -1 + z \right) }\nonumber\\
& & {\bf B}_{21}(z)=  
  \frac{\,\left( 66827 + 122199\,z \right) }{576\,\left( -1 + z \right) }\nonumber\\
& & {\bf B}_{22}(z)=  
  \frac{\,\left( -12139 + 17963\,z \right) }{288\,\left( -1 + z \right) }\nonumber\\
& & {\bf B}_{23}(z)=  
  \frac{\left( 19 + 23\,z \right) }{2\,\left( -1 + z \right) }\nonumber\\
\end{eqnarray}
\section{The Expressions for ${\bf B}_i(t)'s$}
In this appendix, we give the expressions for ${\bf B}_i(t)$'s obtained using Mathematica
that figure in the Picard-Fuchs
equation (\ref{eq:PFdiffeq1}).
\begin{eqnarray}
& & {\bf B}_{23}(t)=( -276 + \frac{( 23 + 19e^t ) t}{2( -1 + e^t ) } )  +\nonumber\\ 
& & {\bf B}_{22}(t)=( 35926 - \frac{253( 23 + 19e^t ) t}{2( -1 + e^t ) } + 
         \frac{( -17963 + 12139e^t ) t^2}{288( -1 + e^t ) } )  + \nonumber\\
& & {\bf B}_{21}(t)=( -2932776 + \frac{30107( 23 + 19e^t ) t}{2( -1 + e^t ) } - 
         \frac{77( -17963 + 12139e^t ) t^2}{96( -1 + e^t ) } + 
         \frac{( 122199 + 66827e^t ) t^3}{576( -1 + e^t ) } )  + \nonumber\\
& & {\bf B}_{20}(t)=( 168423871 - \frac{2240315( 23 + 19e^t ) t}{2( -1 + e^t ) } + 
         \frac{25025( -17963 + 12139e^t ) t^2}{288( -1 + e^t )}-\nonumber\\
& &          {{35( 122199 + 66827e^t ) t^3}\over{96( -1 + e^t ) }} +
\nonumber\\
& &  
{( -168423871 + 73644729e^t ) t^4\over{331776( -1 + e^t ) }}
)\nonumber\\
& &  {\bf B}_{19}(t)=( -7234669596 + \frac{58448313( 23 + 19e^t ) t}{-1 + e^t} - 
         \frac{563255( -17963 + 12139e^t ) t^2}{96( -1 + e^t ) } + \nonumber\\
& &         \frac{20615( 122199 + 66827e^t ) t^3}{576( -1 + e^t ) } - 
         \frac{95( -168423871 + 73644729e^t ) t^4}{165888( -1 + e^t ) } + 
          \frac{361( 1670053 + 576063e^t ) t^5}{663552( -1 + e^t ) } )  + \nonumber\\
& & {\bf B}_{18}(t)=( 241276443496 - \frac{2273023599( 23 + 19e^t ) t}{-1 + e^t} + 
         \frac{6643483( -17963 + 12139e^t ) t^2}{24( -1 + e^t ) } - \nonumber\\
& &         \frac{69825( 122199 + 66827e^t ) t^3}{32( -1 + e^t ) } + 
         \frac{5605( -168423871 + 73644729e^t ) t^4}{110592( -1 + e^t ) } - 
\nonumber\\
& &         \frac{6859( 1670053 + 576063e^t ) t^5}{73728( -1 + e^t ) } + 
         \frac{19( -1587345023 + 425415838e^t ) t^6}{23887872( -1 + e^t ) } )  + \nonumber\\
& & .....................................................................\nonumber\\
& &{\bf B}_2(t)=( 96538966652493066240000 - \frac{2074238389667727360000( 23 + 19e^t ) t}{-1 + e^t} +\nonumber\\
& &  
         \frac{646683370764480000( -17963 + 12139e^t ) t^2}{-1 + e^t}
 - \frac{15196090341600000( 122199 + 66827e^t ) t^3}{-1 + e^t} + \nonumber\\
& &          \frac{1300772650275( -168423871 + 73644729e^t ) t^4}{-1 + e^t} -
\nonumber\\ 
& & 
         \frac{24348032073225( 1670053 + 576063e^t ) t^5}{2( -1 + e^t ) }\nonumber\\
& &  + \frac{140122891475( -1587345023 + 425415838e^t ) t^6}{144( -1 + e^t ) }\nonumber\\  
& &    - \frac{426397825( 66672816001 + 13645602646e^t ) t^7}{288( -1 + e^t ) } +\nonumber\\
& &        \frac{209257475( -137272511800831 + 21034217831392e^t ) t^8}{5308416( -1 + e^t ) }
\nonumber\\
& &  - \frac{451117205( 18227621504883 + 2043640448992e^t ) t^9}{31850496( -1 + e^t ) } + \nonumber\\
& &         \frac{5729965( -17350903224352103 + 1385620419100032e^t ) t^{10}}{9172942848( -1 + e^t ) } -\nonumber\\
& &  
\frac{430105( 17223196429221419 + 948851467387776e^t ) t^{11}}{18345885696( -1 + e^t ) } +
\nonumber\\
& &  \frac{418555( -4070384057007569521 + 148836350338617600e^t ) t^{12}}{126806761930752( -1 + e^t ) } - \nonumber\\
& &         \frac{701195( 145095225300764057 + 3357738175843584e^t ) t^{13}}{253613523861504( -1 + e^t ) } + \nonumber\\
& & \frac{7129(-55246113744858429349+761904150306398208e^t)t^{14}}{36520347436056576(-1+e^t ) } -
\nonumber\\ 
& &\frac{761(2788771435559577853+21189773926434816e^t)t^{15}}{8115632763568128(-1+e^t)}+
\nonumber\\
& & 
 \frac{121( -331544607220340363311 + 1246918912966545408e^t ) t^{16}}{7011906707722862592( -1 + e^t ) }\nonumber\\
& &  - \frac{49( 96816543515895571753 + 154174243770593280e^t ) t^{17}}{42071440246337175552( -1 + e^t ) } + \nonumber\\
& & 
\frac{1 37( -88158910649831337151 + 46134676456243200e^t ) t^{18}}{6058287395472553279488( -1 + e^t ) } - \nonumber\\
& &         \frac{275( 1381067323704722393 + 142216445952000e^t ) t^{19}}{12116574790945106558976( -1 + e^t ) } - \nonumber\\
& & 
         \frac{20871128486031016325t^{20}}{48466299163780426235904( -1 + e^t ) }
\nonumber\\
& &  - \frac{160288719172038625t^{21}}{32310866109186950823936( -1 + e^t ) }
 - \nonumber\\
& & \frac{149679187975625t^{22}}{3590096234354105647104( -1 + e^t ) } )\nonumber\\
& &{\bf B}_1(t)=(-25852016738884976640000 + \frac{562000363888803840000( 23 + 19e^t ) t}{-1 + e^t} - \nonumber\\
& &       \frac{177399104762880000( -17963 + 12139e^t ) t^2}{-1 + e^t} 
\nonumber\\
& & + \frac{4223788208640000( 122199 + 66827e^t ) t^3}{-1 + e^t} - 
\frac{366648282000( -168423871 + 73644729e^t ) t^4}{-1 + e^t} \nonumber\\
& & + \frac{3483158679000( 1670053 + 576063e^t ) t^5}{-1 + e^t}\nonumber\\
& &  - \frac{282907625( -1587345023 + 425415838e^t ) t^6}{-1 + e^t} + 
         \frac{875875( 66672816001 + 13645602646e^t ) t^7}{2( -1 + e^t ) } - 
\nonumber\\
& & 
\frac{875875( -137272511800831 + 21034217831392e^t ) t^8}{73728( -1 + e^t ) } + \nonumber\\
& &         \frac{1926925( 18227621504883 + 2043640448992e^t ) t^9}{442368( -1 + e^t ) } -\nonumber\\ 
& & 
         \frac{25025( -17350903224352103 + 1385620419100032e^t ) t^{10}}{127401984( -1 + e^t ) } + \nonumber\\
& & \frac{1925( 17223196429221419 + 948851467387776e^t ) t^{11}}{254803968( -1 + e^t ) } - \nonumber\\
& &         \frac{1925( -4070384057007569521 + 148836350338617600e^t ) t^{12}}{1761205026816( -1 + e^t ) } + \nonumber\\
& &          \frac{3325( 145095225300764057 + 3357738175843584e^t ) t^{13}}{3522410053632( -1 + e^t ) }
\nonumber\\
& & - \frac{35( -55246113744858429349 + 761904150306398208e^t ) t^{14}}{507227047723008( -1 + e^t ) } + \nonumber\\
& &         \frac{35( 2788771435559577853 + 21189773926434816e^t ) t^{15}}{1014454095446016( -1 + e^t ) } - \nonumber\\
& &          \frac{35( -331544607220340363311 + 1246918912966545408e^t ) t^{16}}{5258930030792146944( -1 + e^t ) } + 
\nonumber\\
& & \frac{5( 96816543515895571753 + 154174243770593280e^t ) t^{17}}{10517860061584293888( -1 + e^t ) } - \nonumber\\
& &          \frac{5( -88158910649831337151 + 46134676456243200e^t ) t^{18}}{504857282956046106624( -1 + e^t ) } +\nonumber\\ 
& & 
         \frac{11( 1381067323704722393 + 142216445952000e^t ) t^{19}}{1009714565912092213248( -1 + e^t ) } + 
\nonumber\\
& &     
     \frac{1897375316911910575t^{20}}{8077716527296737705984( -1 + e^t ) } + 
\frac{160288719172038625t^{21}}{48466299163780426235904( -1 + e^t ) } + 
 \nonumber\\
& &         \frac{149679187975625t^{22}}{3590096234354105647104( -1 + e^t ) }
 + 
         \frac{2505147019375t^{23}}{5385144351531158470656( -1 + e^t ) } ) 
\end{eqnarray}

\end{document}